\documentclass[preprint,fleqn,pre,twocolumn,10pt,showpacs,balancelastpage]{revtex4}
\usepackage{amssymb}

\usepackage{amsmath}
\usepackage{graphicx}

\input{tcilatex}

\begin{document}

\title{Nonlinear dynamics and chaos in parametric sound generation}
\author{V\'{\i}ctor J. S\'{a}nchez-Morcillo}
\author{V\'{\i}ctor Espinosa}
\author{Javier Redondo}
\author{Jes\'{u}s Alba}
\affiliation{Departament de F\'{\i}sica Aplicada, Universitat Polit\'{e}cnica de Val\`{e}%
ncia, Crta. Natzaret-Oliva s/n, 46730 Grau de Gandia, Spain}
\author{Short title: Chaos in acoustic interferometers}
\title{Nonlinear dynamics and chaos in parametric sound generation}
\author{V. J. S\'{a}nchez--Morcillo, V. Espinosa, J. Redondo, and J. Alba}
\affiliation{Departament de F\'{\i}sica Aplicada, Universitat Polit\`{e}cnica de Val\`{e}%
ncia, Crta. Natzaret-Oliva s/n, 46730 Grau de Gandia, Spain}

\begin{abstract}
A theoretical analysis of the subharmonic generation process in an
acoustical resonator (interferometer) with plane walls is performed. It is
shown that, when both the pumping wave and the generated subharmonic are
detuned with respect to the resonator modes, the fields can display complex
temporal behaviour such as self-pulsing and chaos. A discussion about the
acoustical parameters required for the experimental observation of the
phenomenon is given.
\end{abstract}

\maketitle

\section*{I. INTRODUCTION.}

Nonlinear dynamical phenomena have been reported in several acoustical
systems. Different complex scenarios of temporal evolution, including
self-pulsing, period doubling and deterministic chaos, are typically
observed in a large variety of driven oscillators, such as nonlinear
resonators, musical instruments or ultrasonic cavitation (for a review, see
Ref. 1). In the particular resonator configuration, self-action of sound
propagating in a viscid fliud, where the thermal mechanism of nonlinearity
is dominant, has been studied theoretical and experimentally, and the onset
of bistable, self-pulsed and chaotic regimes has been shown to occur with
precedence to another acoustical nonlinear effects related to the elastic
(quadratic) nonlinearity of the medium.\cite{Lyakhov93} These nonthermal
effects, involving the appearance of new frequencies not present in the
driving source, provoke the excitation of higher harmonics of the driving
source (which in the absence of dispersion leads to waveform distortion and
ultimately to the development of shock waves) or the frequency division of
the input wave (subharmonics), an effect so called parametric sound
generation. The latter phenomenon consist in the resonant interaction of a
triad of waves with frequencies $\omega _{0},\omega _{1}$ and $\omega _{2}$,
which are related by the energy conservation condition $\omega _{0}=\omega
_{1}+\omega _{2}$. The process is initiated by an input pumping wave of
frequency $\omega _{0}$ which, due to the propagation in the nonlinear
medium, generates a pair of waves with frequencies $\omega _{1}$ and $\omega
_{2}$. When the wave interaction occurs in a resonator, a threshold value
for the input amplitude is required. This process has been described before
by several authors under different conditions, either theoretical and
experimentally.\cite{Korpel65,Adler70,Yen75,Ostrovsky76}

It is important to note that the presence of higher harmonics can be avoided
by different means. One method considers that the interaction takes place in
a dispersive system. Dispersion allows that only few modes, those satisfying
given synchronism conditions, participate in the interaction process. In
finite geometries, such as waveguides \cite{Hamilton87} or resonators,\cite%
{Ostrovsky78} the dispersion is introduced by the lateral boundaries.
Different resonance modes, propagating at different angles, propagate with
different effective\ phase velocities. However, in homogeneous unbounded
systems dispersion is usually not present. Different dispersion mechanisms
have been proposed in nonlinear acoustics, such as propagation in bubbly or
layered (periodic) media,\cite{Hamiltonbook} which makes sound velocity
propagation to be wavelength dependent. Other proposed mechanisms are, for
example, media with selective absorption, in which selected spectral
components experience strong losses and may be removed from the wave field.%
\cite{Zarembo74}, or resonators where the end walls present some
frequency-dependent complex impedance.\cite{Yen75} In this case, the
resonance modes of the resonator are not integrally related, and by proper
adjustment of the resonator parameters one can get that only few modes,
those lying close to a cavity resonance, reach a signifficant amplitude.
Under this conditions, a spectral approach to the problem results
appropriate, as has been proposed and experimentally confirmed in Ref. 6.

Therefore the selective effect of the resonator allows to reduce the study
of parametric sound generation to the interaction of three field modes,
correspondig to the driving (fundamental) and subharmonic frequencies, and
to describe this interaction through a small set of nonlinear coupled
differential equations. In the present work we describe the particular
degenerate case of subharmonic generation, where $\omega _{1}=\omega _{2}$
and consequently $\omega _{0}=2\omega _{1},$ being $\omega _{0}$ the
fundamental and $\omega _{1}$ generated subharmonic, both quasi-resonant
with a corresponding resonance mode. This degenerate case has been the
matter of previous experimental studies,\cite{Yen75,Ostrovsky78} and the
observation of dynamical behaviour has been reported.\cite%
{Ostrovsky78,cook89} However, despite the experimental observation of
complex temporal dynamics in this system, a theoretical framework supporting
these effects and taking into account the peculiarities of the acoustic
systems is absent. The main purpose of this work is to establish the
necessary conditions for the development of nonlinear dynamics in an
acoustic resonator where parametric generation of sound takes place.

\section*{II. MODEL EQUATIONS.}

We consider a resonator with plane walls with finite width (acoustic
interferometer) driven at a frequency $\omega _{0}$, and that the wave
interaction is only effective among two resonant frequencies: the pump $%
\omega _{0}$ and the subharmonic $\omega _{1}$, for which the relation $%
\omega _{0}=2\omega _{1}$ hold. The field inside the resonator can then be
expanded as 
\begin{equation}
p(\mathbf{r},t)=\sum_{j=1}^{2}P_{j}(\mathbf{r},t),  \label{sum}
\end{equation}%
where $P_{j}(\mathbf{r},t)$ are the wave components related with the
frequency $\omega _{j}$. Restricting the analysis to a one-dimensional
resonator, where quasi-plane waves propagate on the $z$ axis, and taking
into account that, due to reflections in the walls, there exist waves
propagating simultaneously in opposite directions, the wave components take
the form%
\begin{equation}
P_{j}(z,t)=p_{j}(t)\cos \left( k_{j}^{c}z\right) e^{-i\omega _{j}t}+c.c.,
\label{cosmodes}
\end{equation}%
where $k_{j}^{c}$ a cavity eigenmode.

Under the assumption of plane waves inside a resonator with highly
reflecting ends (weak cavity losses), the model for the time evolution of
degenerate parametric sound generation reads \cite{victorsm03}%
\begin{align}
\frac{\partial p_{0}}{\partial t}& =E-\gamma _{0}(1+i\Delta
_{0})p_{0}-i\beta p_{1}^{2},  \notag \\
\frac{\partial p_{1}}{\partial t}& =-\gamma _{1}(1+i\Delta _{1})p_{1}-i\beta
p_{1}^{\star }p_{0},  \label{model}
\end{align}%
where the variables $p_{i}\;(i=0,1)$ are the complex slow-varying amplitudes
of the fundamental and subharmonic fields, with angular frequencies $\omega
_{i}.$ The rest of parameters and their units are: the pump $E$ ($Pa/s$),
proportional to the injected amplitude $p_{in}$ ($Pa$) of the fundamental
mode and acting as a control parameter;\ the loss coefficient $\gamma _{i}$ (%
$s^{-1}$); the nonlinear coupling parameter $\beta $ ($s^{-1}Pa^{-1}$)$,$
and the adimensional detuning parameter $\Delta _{i}$, proportional to the
frequency mistuning between\ the field frequency $\omega _{i}$\ and the
closest resonator eigenfrequencies $\omega _{i}^{c}$. All these parameters
are defined as%
\begin{align}
E& =\frac{c}{2L}\sqrt{\mathcal{T}_{0}}p_{in},  \notag \\
\gamma _{i}& =\frac{c\mathcal{T}_{0}}{2L}+\kappa _{i},  \notag \\
\Delta _{i}& =\frac{\omega _{i}^{c}-\omega _{i}}{\gamma i},
\label{parameters} \\
\beta & =\frac{\varepsilon \omega _{1}}{4\rho _{0}c^{2}}.  \notag
\end{align}%
being $c$ the propagation velocities, $\mathcal{T}_{i}\;$and $\mathcal{R}%
_{i},\ $the transmission and reflection coefficients at resonator ends ($%
\mathcal{R}_{i}>>\mathcal{T}_{i}$ has been assumed), $L$ the length of the
resonator, $\varepsilon \;$the nonlinearity parameter, $\rho _{0}$ the
medium density and $\kappa _{i}=\delta k_{i}^{2}/2$ account for losses
related to diffusivity of sound in the medium $\delta $. A detailed
derivation of this model, generalized to include the field dependence on the
transversal coordinates (not considered here for simplicity) can be found in
Ref. 15. The same equations have been obtained in Ref. 3 for waveguide
resonators (althought with different coupling coefficients, dependent on the
structure of the interacting transverse modes), or for one-dimensional
acoustic interferometers in Ref. 6, in a form identical to Eqs. (\ref{model}%
).

The model (\ref{model}) has been also derived in the context of nonlinear
optics.\cite{Lugiato88} where the existence of complex temporal dynamics has
been reported. In the next section we review the main results concerning the
solutions and its stability properties for our particular acoustic system.

\section*{III. STATIONARY SOLUTIONS.}

Two stationary solutions of (\ref{model}) are obtained when the temporal
derivatives vanish.\cite{Ostrovsky76} The simplest case corresponds to the
trivial solution, 
\begin{equation}
\overline{p}_{0}=\frac{E}{\gamma _{0}\left( 1+i\Delta _{0}\right) },\,%
\overline{p}_{1}=0,  \label{trivial}
\end{equation}%
characterized by a null value of the subharmonic field inside the resonator.
Increasing the pump value $E$ this solution bifurcates into a new solution,
in which also the subharmonic field has a nonzero amplitude $\left\vert 
\overline{p}_{1}\right\vert $, given by 
\begin{eqnarray}
\left\vert \overline{p}_{1}\right\vert ^{2} &=&\frac{1}{\beta ^{2}}(\gamma
_{0}\gamma _{1}(\Delta _{0}\Delta _{1}-1)\pm  \label{notriv1} \\
&&\sqrt{\beta ^{2}\left\vert E\right\vert ^{2}-\gamma _{0}^{2}\gamma
_{1}^{2}\left( \Delta _{0}+\Delta _{1}\right) ^{2}}),  \notag
\end{eqnarray}%
while the stationary fundamental amplitude $\left\vert \overline{p}%
_{0}\right\vert $ takes the value 
\begin{equation}
\left\vert \overline{p}_{0}\right\vert =\frac{\gamma _{1}}{\beta }\sqrt{%
1+\Delta _{1}^{2}},  \label{notriv0}
\end{equation}

The emergence of this finite amplitude solution corresponds to the process
of subharmonic generation. Note that the fundamental amplitude above the
threshold is independent of the value of the injected pump, which means that
all the energy is transferred to the subharmonic wave.

The transition from solutions (\ref{trivial}) to (\ref{notriv1}) occurs at a
critical pump amplitude%
\begin{equation}
\left\vert E\right\vert =\frac{\gamma _{0}\gamma _{1}}{\beta }\sqrt{\left(
1+\Delta _{0}^{2}\right) \left( 1+\Delta _{1}^{2}\right) },
\label{1erumbral}
\end{equation}%
beyond which the trivial solution is unstable. These results have been
confirmed experimentally in Ref. 3. The character of the bifurcation depends
on the detuning values. As demonstrated in Ref. 3, and also in the optical
context,\cite{Lugiato88} the bifurcation is supercritical when $\Delta
_{0}\Delta _{1}<1$, and subcritical when $\Delta _{0}\Delta _{1}>1$. In the
latter case, both trivial and finite amplitude solutions can coexist for
given sets of the parameters, which results in a regime of bistability
between different solutions. Figure 1 illustrates the form of the solutions
in these two regimes.

\begin{figure}[h]
\centering\includegraphics[width=7cm]{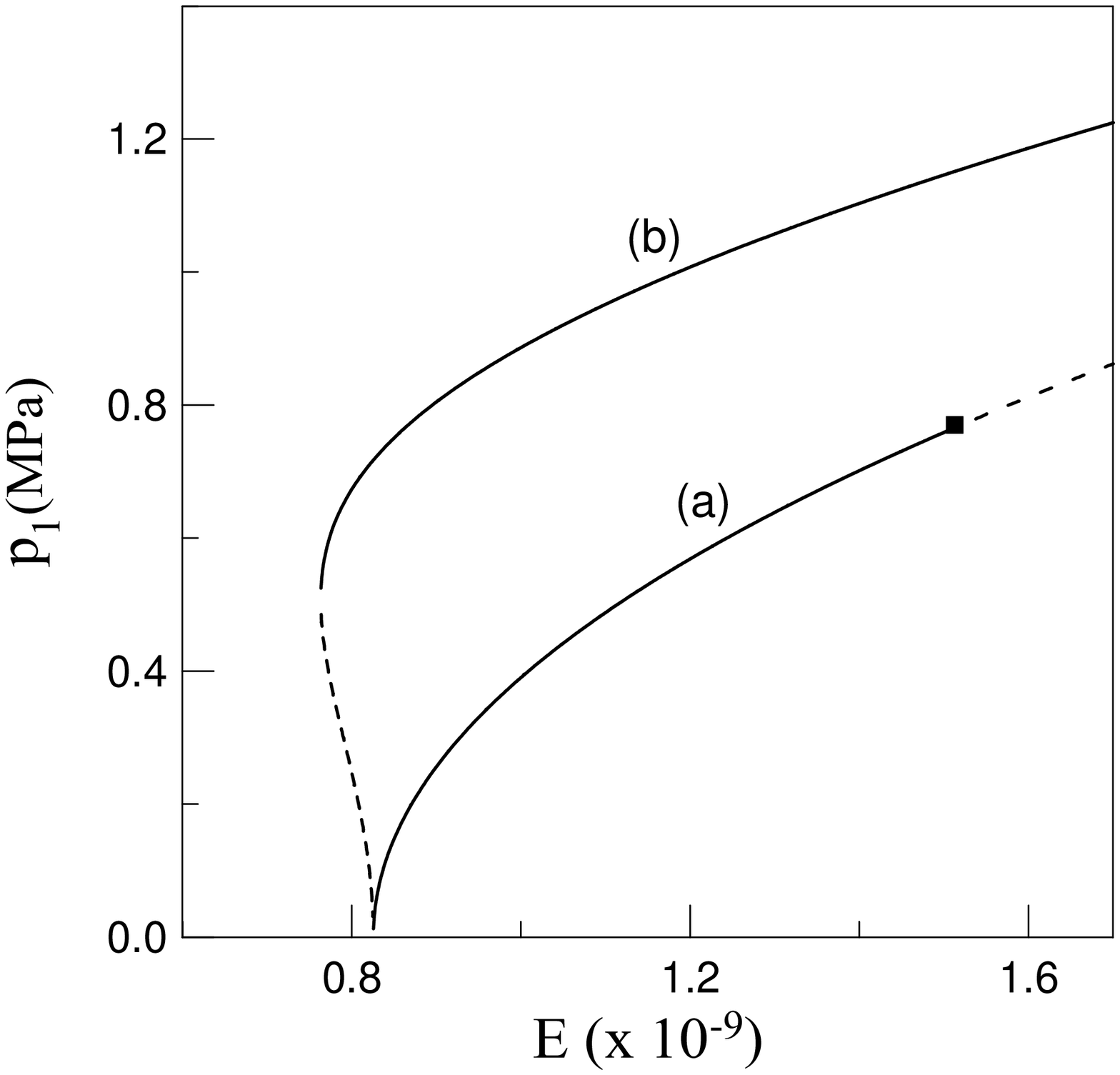}
\caption{Bifurcation diagram of the subharmonic field in the monostable (a)
and bistable (b) cases. Subharmonic detuning values are $\Delta _{1}=2.4$
(a), and $\Delta _{1}=-2.4$ (b). The rest of parameters are $\Delta _{0}=1,$ 
$\protect\beta =0.002$ and $\protect\gamma _{0}=\protect\gamma _{1}=500$
(see text below)$.$ Pressure units are given in $Pa$. Unstable branches are
depicted in dashed lines, and the black square correspond to a Hopf
bifurcation point. }
\end{figure}

\section*{IV. LINEAR STABILITY ANALYSIS.}

In order to demonstrate the possibility of temporal dynamics in this system,
the stability of the previous stationary solutions has been investigated by
substituting in Eqs. (\ref{model}) and their complex conjugate a perturbed
solution in the form 
\begin{equation}
p_{i}\left( t\right) =\overline{p}_{i}+\delta p_{i}(t)=\overline{p}_{i}+%
\overline{\delta p}_{i}\;e^{\lambda t},  \label{deviation}
\end{equation}%
where $\overline{p}_{i}$ are the particular stationary values. By
linearizing the resulting equations around the small perturbations $\delta
p_{i}$, one obtains $\lambda $ as the eigenvalues of the stability matrix
that relates the vector of the perturbations $\left( \delta p_{i}(t),\delta
p_{i}^{\ast }(t)\right) $ with their temporal derivatives. The instability
of the solution is determined by the existence of positive real parts of the
roots $\lambda $ of the\ characteristic polynomial,%
\begin{equation}
\lambda ^{4}+c_{1}\lambda ^{3}+c_{2}\lambda ^{2}+c_{3}\lambda +c_{4}=0
\label{charpol1}
\end{equation}%
where, for the finite amplitude solution Eq. (\ref{notriv1}) and (\ref%
{notriv0}), the coefficients take the form%
\begin{align}
c_{1}& =2\left( \gamma _{0}+\gamma _{1}\right) ,  \notag \\
c_{2}& =4\beta ^{2}\left\vert \overline{p}_{1}\right\vert ^{2}+\gamma
_{0}^{2}+4\gamma _{0}\gamma _{1}+\gamma _{0}\Delta _{0}^{2},  \notag \\
c_{3}& =4\beta ^{2}\left\vert \overline{p}_{1}\right\vert ^{2}\left( \gamma
_{0}+\gamma _{1}\right) +2\gamma _{0}^{2}\gamma _{1}\left( 1+\Delta
_{0}^{2}\right) ,  \notag \\
c_{4}& =4\beta ^{2}\left\vert \overline{p}_{1}\right\vert ^{2}\left( \beta
^{2}\left\vert \overline{p}_{1}\right\vert ^{2}-\gamma _{0}\gamma _{1}\left(
\Delta _{0}\Delta _{1}-1\right) \right) .  \label{charpol2}
\end{align}

Instead of solving the quartic polynomial Eq. (\ref{charpol1}) to find the
eigenvalues, we apply the Hurwitz criterion (see e.g. Ref. 17), which states
that the solution becomes unstable when at least one of the following
conditions are violated: 
\begin{subequations}
\begin{align}
& c_{1}\eqslantgtr 0,\ c_{2}\eqslantgtr 0,\ c_{3}\eqslantgtr 0,\
c_{4}\eqslantgtr 0  \label{hurwitz1} \\
& c_{1}c_{2}-c_{3}\eqslantgtr 0  \label{hurwitz2} \\
& c_{3}\left( c_{1}c_{2}-c_{3}\right) -c_{1}^{2}c_{4}\eqslantgtr 0.
\label{hurwitz3}
\end{align}

Condition (\ref{hurwitz2}) is always fulfilled. In conditions (\ref{hurwitz1}%
), only $c_{4}\;$can become negative, corresponding with the turning point
of the stationary solution of the $\Delta _{0}\Delta _{1}>1\;$case (see Fig.
1b). This means that the lower branch of the stationary solution,
represented by the dashed line in Fig. 1 is always unstable.

Finally, the equality in condition (\ref{hurwitz3}) denotes the existence of
a pair of complex conjugate eigenvalues with vanishing real part ($\lambda
=\pm i\omega $), which are indicative of a Hopf bifurcation which gives rise
to oscillating solutions, with angular frequency $\omega $. Substituting (%
\ref{charpol2}) in (\ref{hurwitz3}) the instability condition reads 
\end{subequations}
\begin{eqnarray}
&&\gamma _{0}^{2}\gamma _{1}\left( 1+\Delta _{0}^{2}\right) \left( \left(
\gamma _{0}+2\gamma _{1}\right) ^{2}+\gamma _{0}^{2}\Delta _{0}^{2}\right) +
\label{Hopfcond} \\
&&2\left( \gamma _{0}+\gamma _{1}\right) ^{2}\left( \gamma _{0}\left(
1+\Delta _{0}^{2}\right) +2\gamma _{1}\left( 1+\Delta _{0}\Delta _{1}\right)
\right) \beta ^{2}\left\vert \overline{p}_{1}\right\vert ^{2}\geqq 0.  \notag
\end{eqnarray}%
which is fullfilled when the subharmonic amplitude reaches the value%
\begin{equation}
\left\vert \overline{p}_{1}\right\vert _{HB}^{2}=\frac{-\gamma
_{0}^{2}\gamma _{1}\left( 1+\Delta _{0}^{2}\right) \left( \left( \gamma
_{0}+2\gamma _{1}\right) ^{2}+\gamma _{0}^{2}\Delta _{0}^{2}\right) }{2\beta
^{2}\left( \gamma _{0}+\gamma _{1}\right) ^{2}\left( \gamma _{0}\left(
1+\Delta _{0}^{2}\right) +2\gamma _{1}\left( 1+\Delta _{0}\Delta _{1}\right)
\right) },  \label{int1hopf}
\end{equation}%
The positiveness of the threshold value (\ref{int1hopf}) requires an
additional condition, namely that the product of the detunings satisfies%
\begin{equation}
\Delta _{0}\Delta _{1}<-\frac{1}{2\gamma _{1}}\left( \gamma _{0}+2\gamma
_{1}+\gamma _{0}\Delta _{0}^{2}\right) .  \label{negatdetuning}
\end{equation}

The latter condition excludes bistability (which requires $\Delta _{0}\Delta
_{1}>1$). Consequently, condition (\ref{negatdetuning}) implies that the
bifurcation that leads to subharmonic generation is always supercritical,
and then only single-valued solutions can undergo a Hopf bifurcation (see
Fig. 1).\ The angular frequency of the oscillations at the Hopf instability
point can be found by substituting $\lambda =i\omega $ in (\ref{Hopfcond}),
and reads 
\begin{equation}
\omega _{HB}=\sqrt{2\beta ^{2}\left\vert p_{1}\right\vert _{HB}^{2}+\frac{%
\gamma _{0}^{2}\gamma _{1}\left( 1+\Delta _{0}^{2}\right) }{\gamma
_{0}+\gamma _{1}}},  \label{freqHB}
\end{equation}%
where $\left\vert p_{1}\right\vert _{HB}$\ is the subharmonic amplitude at
the Hopf bifurcation point given by Eq. (\ref{int1hopf}).

\section*{V. ACOUSTICAL ESTIMATES.}

For a better understanding of the predictions of the model and to assure
their experimental validity, a special care must be taken when choosing the
acoustical parameters involved in the theory. We will first refer such
quantities to the experimental investigation of subharmonic generation in an
acoustic interferometer described in detail by Yen \cite{Yen75}. It consists
in a three sectioned resonator (quartz crystal$\mathcal{-}$water$\mathcal{-}$%
quartz crystal) of variable length, operated at 26$%
{{}^\circ}%
$C. Our aim is to evaluate the threshold for subharmonic generation Eq. (\ref%
{1erumbral}) and the possibility of its destabilization at the value given
by (\ref{int1hopf}). From Ref. 6 we obtain realistic acoustical values for
the field decay rates $\gamma _{i}$, the detunings $\Delta _{i}$ and the
nonlinearity parameter $\beta $ involved in the model. These parameters
depend, as defined in (\ref{parameters}), on sound frequencies and
velocities, medium nonlinearity $\varepsilon $, cavity length and
reflectivity of boundaries, and the eigenfrequencies of the resonator. In
Ref. 6 the resonance modes of the one-dimensional resonator were calculated
by using the formula%
\begin{eqnarray}
&&R\left( \tan k_{w}^{\prime }D+\tan k_{w}^{\prime }H\right) +\tan k^{\prime
}S-  \label{modes} \\
&&R^{2}\tan k_{w}^{\prime }D\tan k_{w}^{\prime }H\tan k^{\prime }S=0  \notag
\end{eqnarray}%
where $R$ is the ratio of wall impedance to medium impedance, $k^{\prime }$
and $k_{w}^{\prime }$ are the wavenumbers in the fluid and the walls
respectively, and $D,H$ and $S$ are the thicknesess of the two end walls and
the length of the actual resonator, respectively. From (\ref{modes}) results
an non-equidistant spectrum, which is required to allow the fulfillment of
the condition (\ref{negatdetuning}). In the case of an equidistant spectrum
(corresponding to a perfect resonator without losses), the detunings are
obeyed to have the same sign.

The values of these resonance modes, as well as the quality factor $Q_{i}$\
associated with them, were also experimentally measured in Ref. 6. The
pumping frequency was about 1.6 MHz, for which a subharmonic emission of 0.8
MHz is expected. The closest resonator eigenfrequencies to such fundamental
and subharmonic emissions were $f_{0}^{c}=1.5853$ MHz and $f_{1}^{c}=0.8088$
MHz, with associated quality factors of 4700 and 2600, respectively. The
quality factor allows to estimate directly the decay rates $\gamma _{i}$
using the relation $Q_{i}=\omega _{i}/2\gamma _{i}$, and the corresponding
detuning is calculated then from (\ref{parameters}). The nonlinear coupling
parameter\ $\beta $ defined in Eq. (\ref{parameters}) is determined by the
subharmonic frequency $\omega _{1},$ the medium nonlinearity ($\varepsilon
=3.5$ for water), and the sound velocity $c=1500\ $m/s. Note that the
possible values of the detunings $\Delta _{0}$ and $\Delta _{1}$ are
restricted by the non-equidistant eigenfrequencies of the resonator and the
parametric character of the process, since $\omega _{0}=2\omega _{1}$.

The threshold of subharmonic generation was also measured by Yen,\cite{Yen75}
and given in terms of the fundamental pressure amplitude inside the
resonator at the bifurcation\ point given by Eq. (\ref{notriv0}). The model
predicts its dependence on the subharmonic field detuning and on the
subharmonic frequency through the coupling parameter $\beta $. The lowest
theoretical value of the threshold corresponds to a perfect tuning of the
subharmonic field with a cavity eigenmode, that is for a fixed $f_{1}=0.8088$
MHz and the corresponding fundamental $f_{0}=1.6176$ MHz, results $\Delta
_{1}=0,$ and $\Delta _{0}=-187.7$, and nonlinear coupling parameter $\beta
=0.002$ $s^{-1}Pa^{-1}$. The predicted fundamental pressure amplitude at
threshold is therefore about $\left\vert p_{0}\right\vert =0.49$ MPa or $4.9$
bar. The lowest threshold reported in Ref. 6 is about $8.8$ bar, measured
for a frequency close to the cavity mode $f_{0}^{c}=1.5853$ MHz. This
difference can be due to the finite bandwidth of the exciting transducer in
opposition to the monochromatic pump considered here, since the introduction
of small detunings on the expected subharmonic field causes a sensitive and
rapid increase of the subharmonic emission threshold.

After the switching-on of the subharmonic field the system supports constant
pressure amplitudes of both fundamental and subharmonic fields,
corresponding to the solutions (\ref{notriv1}) and (\ref{notriv0}). In order
to observe self-oscillations of the amplitudes, the Hopf bifurcation
condition (\ref{negatdetuning}) must be fulfilled, e.g. by increasing the
detuning of the subharmonic field. One would get this by tuning the
frequency of the subharmonic up to the value $f_{1}=0.8072\;$MHz, for which $%
\Delta _{1}=52.4551$ and $\Delta _{0}=-94.7686$. Unfortunately, when trying
to achieve the detunings condition for instabilities, the subharmonic
emission threshold grows up to the unrealistic value $\left\vert \overline{p}%
_{0}\right\vert =26$ MPa. For the same parameters, the Hopf bifurcation
threshold is obtained for a subharmonic amplitude of $\left\vert \overline{p}%
_{1HB}\right\vert =1.14$ GPa. Clearly, for this particular resonator mode
distribution the operation regime should be always in the steady state with
constant pressure amplitudes. This could be the reason why dynamical states
have not been described neither in Ref. 6 nor in other experimental works in
which the conditions were probably similar.

Nevertheless the destabilisation of the subharmonic regime would explain
certain phenomena previously described in the literature, such as
self-modulation of the fields \cite{Ostrovsky78} or possible chaotic
behaviour \cite{cook89}. We estimate next the necessary conditions for the
occurrence of these phenomena at reasonable values. The resonator described
in Ref. 11 corresponds to that described above; two air backed quartz
transducers placed in a Fabry-Perot configuration with water, with a
regulable separation between 5 and 10 cm. The driven transducer had a
resonant frequency $f_{0}=$ $1$ MHz and the other had a lower resonance
frequency. We take this frequency as the fundamental for our estimations and
evaluate the remaining parameters of the desired resonator.\ From inspection
of Eq. (\ref{1erumbral}) it follows that the subharmonic emission threshold
can be lowered by improving the quality of the resonator for both modes.
However, the behaviour of the second threshold it is not so evident: Fig. 2
shows the dependence of the pressure amplitude of the subharmonic field at
the Hopf bifurcation on the ratio of the quality factors affecting to both
fundamental and subharmonic fields. It evidences that instabilities are
reached earlier when the fundamental field has significant lower losses than
the corresponding subharmonic. According to this, we set $Q_{0}=6000$ and $%
Q_{1}=3000,$ which are realistic values if compared to those measured in
Ref. 6.

\begin{figure}[h]
\centering\includegraphics[width=6.5cm]{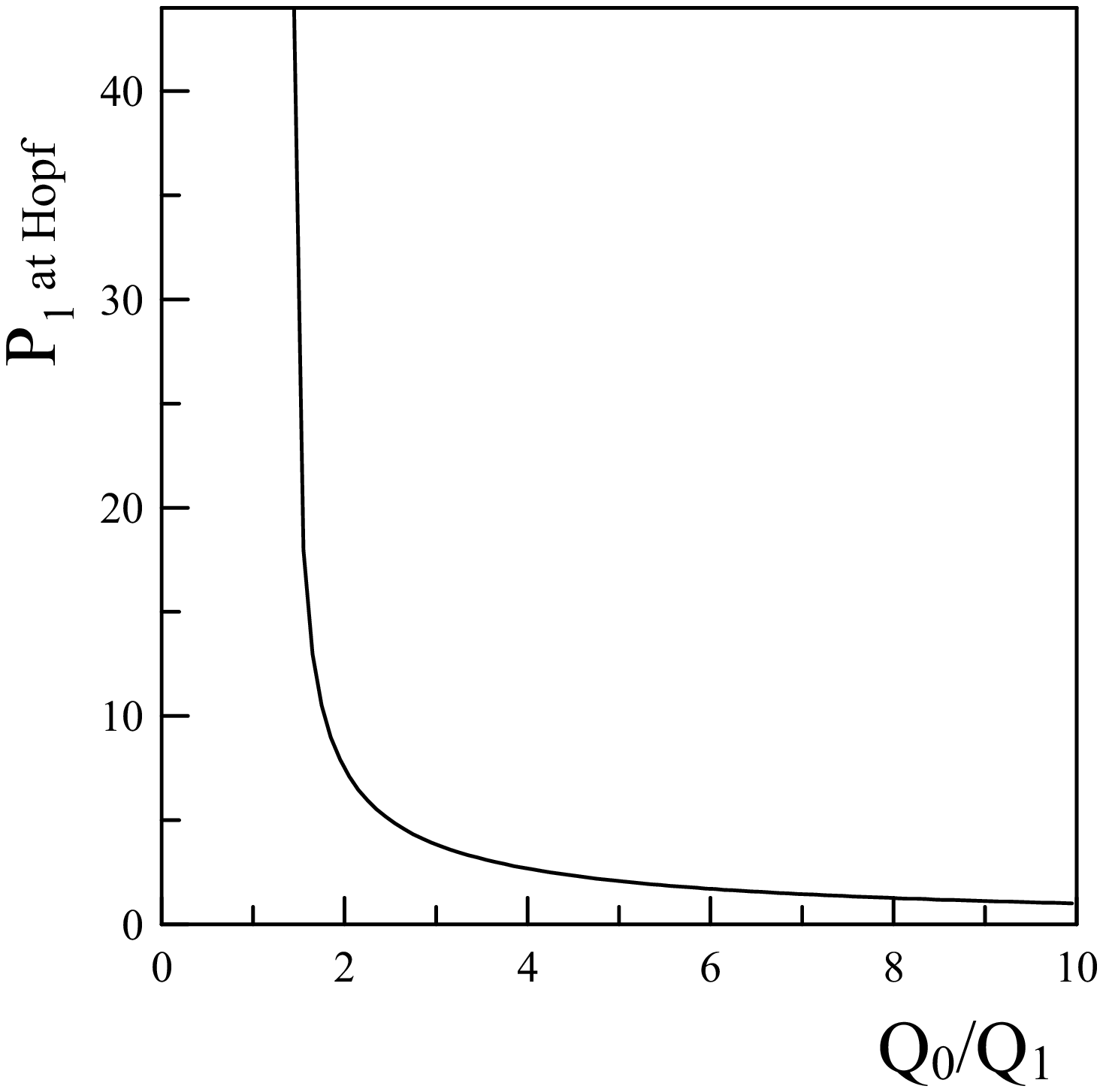}
\caption{Dependence of the pressure amplitude of the subharmonic field at
the Hopf bifurcation ($\left\vert \overline{p}_{1}\right\vert _{HB}$) on the
ratio of the quality factors of fundamental and subharmonic fields. The
parameters are $\Delta _{0}=1,\Delta _{1}=-2.4,\protect\beta =0.002$ and $%
Q_{1}=3000.$ The value of $Q_{0}$ is kept free.}
\end{figure}

The resonator must meet another important restriction concerning its
frequency distribution, in order to satisfy Eq. (\ref{negatdetuning}). In
Fig. 3 are shown the combination of detunings that accomplish this condition
for the given quality factors. The shadowed area represent detuning values
which are compatible with the Hopf bifurcation condition (the inverted
symmetrical figure, with opposite signs of both detunings, is also valid).
One must also consider the lowest possible detunings in order to achieve the
minimal value for the pressure amplitudes in both thresholds. Note that,
whilst $\Delta _{0}$ can be arbitrarily small, the absolute value of $\Delta
_{1}$ must be greater than $2$. A combination of values that reduce the
thresholds to common experimental levels is $(\Delta _{0},\Delta
_{1})=(1.0,-2.4)$, corresponding to an intracavity pressure amplitude of the
fundamental mode at the first threshold of $\left\vert \overline{p}%
_{0}\right\vert =1.11$ MPa, and a subharmonic amplitude pressure at the Hopf
bifurcation point of $\left\vert \overline{p}_{1HB}\right\vert =0.758\ $MPa.
From this discussion it follows that, in order to obtain experimentally the
destabilisation of the subharmonic field, the resonator must be designed
with a mode distribution that allows such values of detunings.

\begin{figure}[h]
\centering\includegraphics[width=6.5cm]{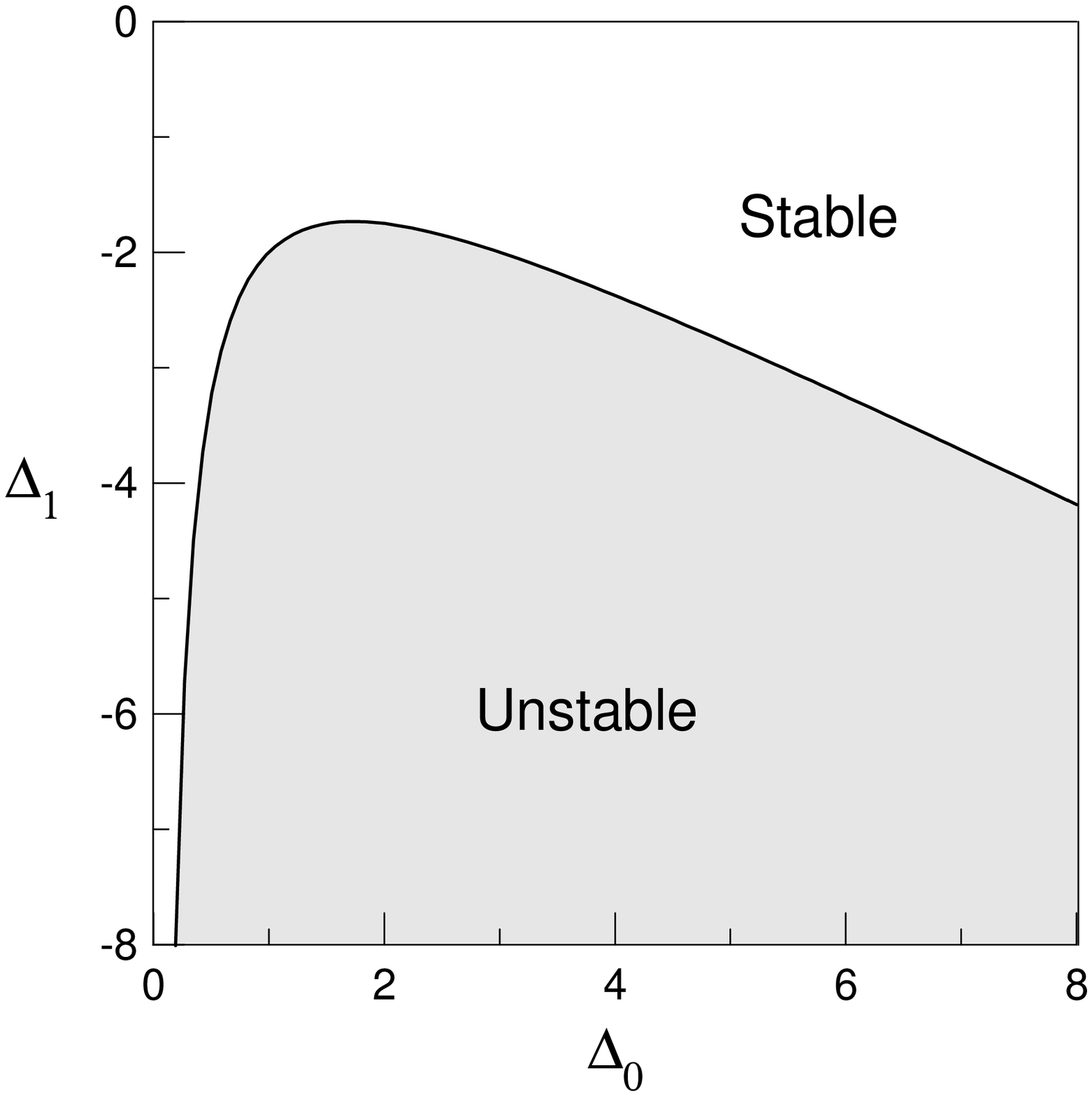}
\caption{Detuning combinations required to fulfil the Hopf bifurcation
condition.}
\end{figure}

The linear stability analysis can not predict the behaviour of the fields
beyond the bifurcation points, and also does not give information about the
subcritical or supercritical character of the bifurcation. In the next
section we perform the numerical integration of the model (\ref{model})
using the parameters estimated above.

\section*{VI. NUMERICAL SIMULATIONS.}

Equations (\ref{model}) have been numerically integrated by means of a
fourth-order Runge-Kutta algorithm, and using the acoustical parameters
discussed in the previous section. According to this, we have chosen the
numerical values $\Delta _{0}=1,$ $\Delta _{1}=-2.4,$ $\gamma _{0}=\gamma
_{1}=523.6$ ($s^{-1}$) and $\beta =0.0012$ ($s^{-1}Pa^{-1}$)$.$ As the
control parameter we consider now the physical variable of the pump
amplitude incident on the resonator $\left\vert p_{in}\right\vert $, related
to the pump value $\left\vert E\right\vert $ appearing in the Eqs. (\ref%
{model}) as $\left\vert p_{in}\right\vert =(2L/c)\sqrt{1/\mathcal{T}}$ $%
\left\vert E\right\vert $. The resonator has a length of $L=10$ cm, and the
end walls a reflectivity coefficient $\mathcal{R}=0.9$, which is in
agreement with the decay rate values considered above.

\begin{figure}[h]
\centering\includegraphics[width=9cm]{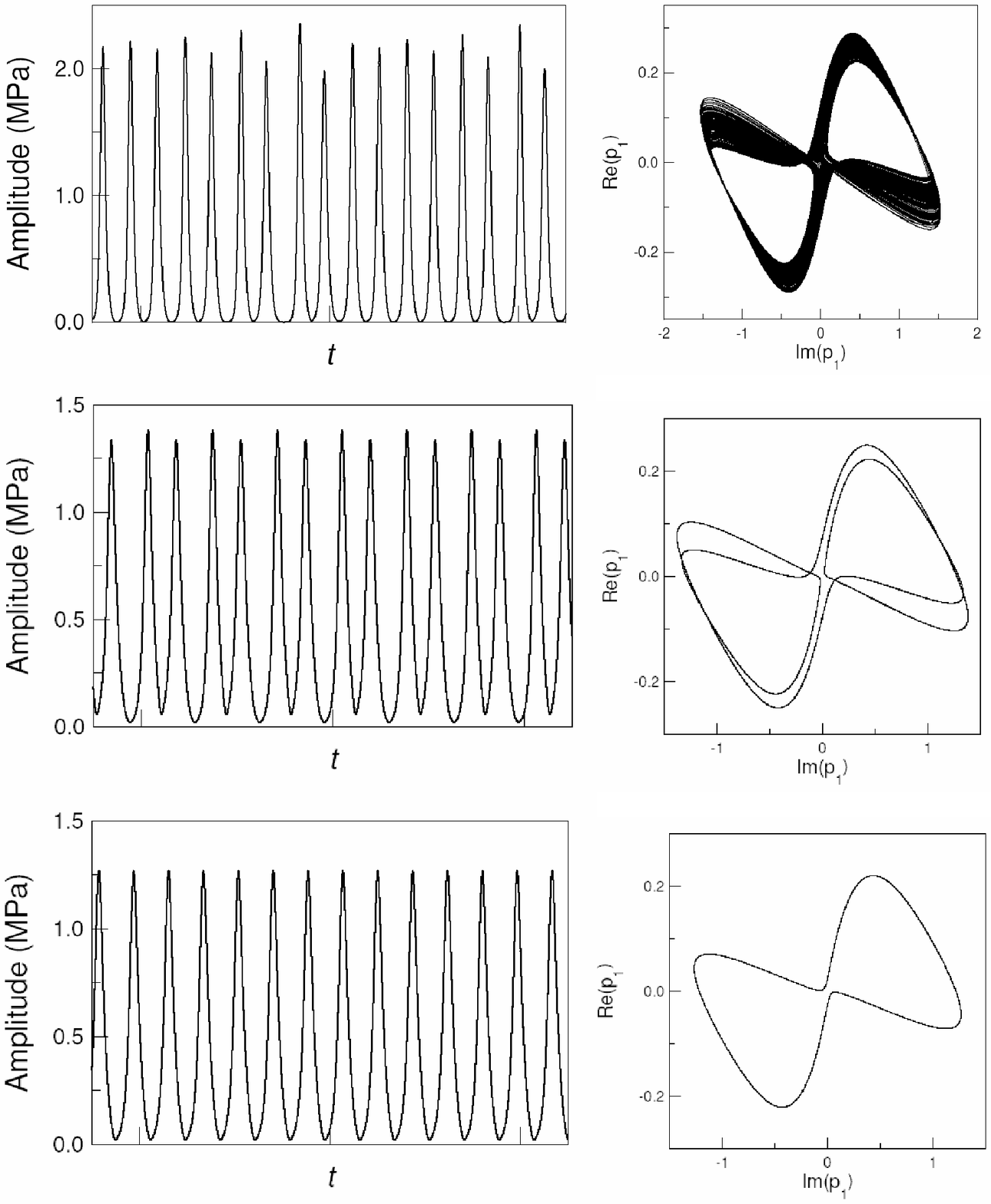}
\caption{ Time evolution (left) and phase portrait (right) of the
sunharmonic field for $\Delta _{0}=1,\Delta _{1}=-2.4,\protect\beta =0.0012, 
$ $\protect\gamma _{0}=\protect\gamma _{1}=523.6,$ and different drivings.
(a) $\left\vert p_{in}\right\vert =0.60$ MPa. (b) $\left\vert
p_{in}\right\vert =0.58$ MPa. (c) $\left\vert p_{in}\right\vert =0.55$ MPa.}
\end{figure}

The numerical results are summarized in Fig. 4. The show the existence of a
bifurcation of the subharmonic solution Eq. (\ref{notriv1}) at the injected
amplitude $\left\vert p_{in}\right\vert =0.60$ MPa, as predicted by the
linear stability analysis. However, the temporal evolution of the fields
near the bifurcation point is not oscillatory but chaotic, as shown in Fig.
4(a). At the right, the corresponding phase portrait is shown, where the
dense structure of the attractor is appreciated, as corresponds to chaotic
behaviour. When decreasing the injected amplitude below the bifurcation
point, the temporal behaviour persists, denoting that the Hopf instability
is subcritical. As pumping is decreased , the temporal evolution of the
amplitude follows one of the universal scenarios leading to chaos, namely a
period-doubling scenario or Feigenbaum route to chaos. Since we are
decreasing the control parameter, we observe the transition from chaos to
periodicity in an inverted way. Slightly below the bifurcation point,
regular oscillations are observed with period eight ($P_{8}$) and period
four ($P_{4}$), which exist in a narrow window. At $\left\vert
p_{in}\right\vert =0.58$ MPa, the evolution is bi-periodic ($P_{2}$) as
shown in Fig. 4(b). For $\left\vert p_{in}\right\vert =0.55$ MPa the
evolution is periodic with a single period ($P_{1}$). This case is
represented in Fig. 4(c). Further decreasing the control parameter, the
system again enters in a chaotic regime, mediated a reversed period doubling
cascade. This chaotic regime exist only in a narrow window, and the scenario
reported above repeats until the value $\left\vert p_{in}\right\vert =0.53$
MPa, where the subharmonic field jumps again to the stable stationary branch
given by Eqs. (\ref{notriv1}). This corresponds to the turning point of the
dynamic branch, associated to a fold or saddle node bifurcation.

The complete scenario of bifurcations can be represented in a single plot\
(Fig. 5), where the frequency spectrum of solutions is given as a function
of the control parameter. The subcritical character of the bifurcation, and
the complexity of the temporal evolution in this system is evident here.
Several chaotic windows are observed (dark horizontal bands, corresponding
to a dense spectrum), always followed by a period-doubling sequence where $%
P_{8}$, $P_{4}$, $P_{2}$ and $P_{1}$ states are observed. In the plot a grey
scale has been used, representing pressure leves ranging from 0 to 100 dB$.$

\begin{figure}[h]
\centering\includegraphics[width=8cm]{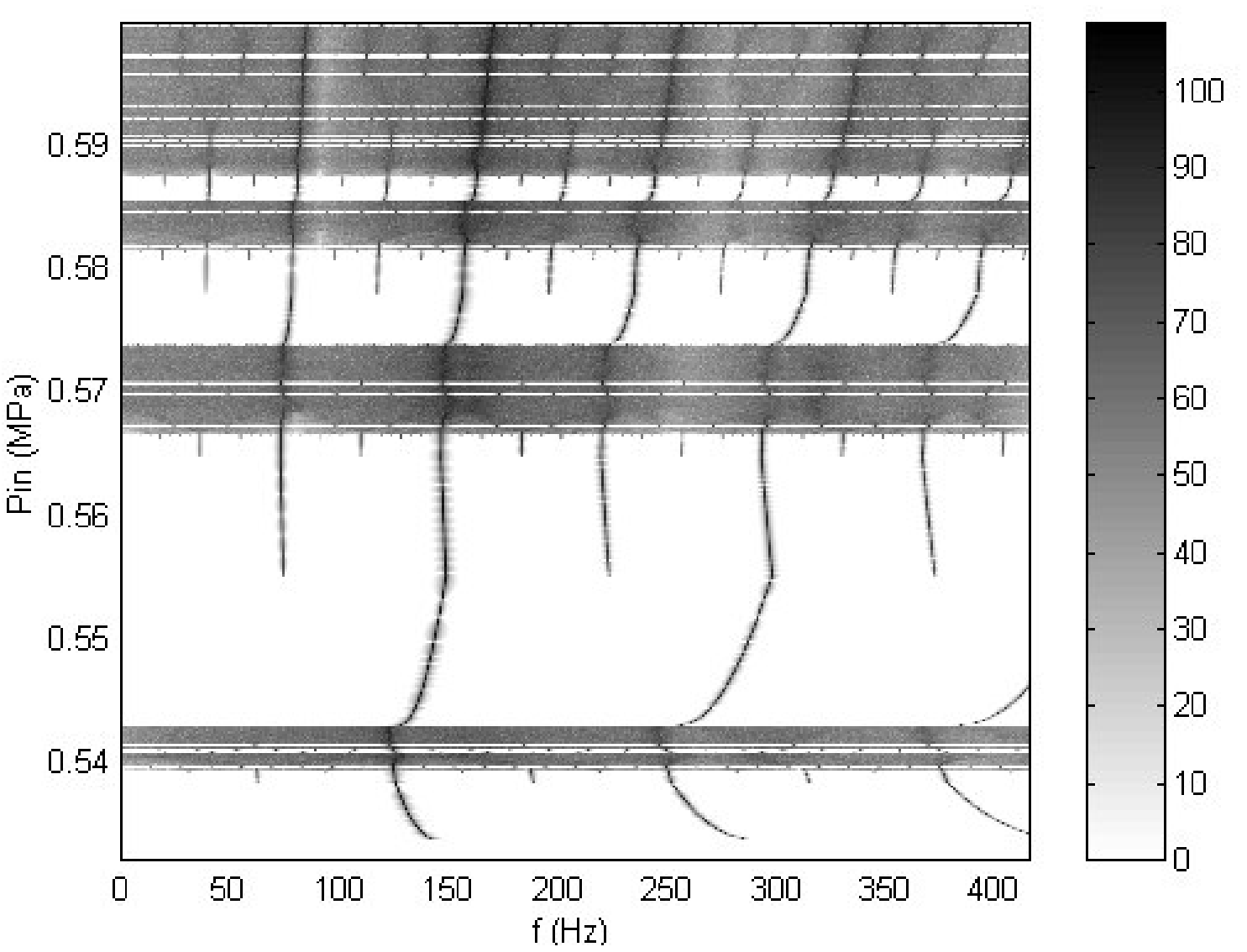}
\caption{Spectrum of solutions as a function of the injected pump amplitude,
for the parameters given in Fig. 4.}
\end{figure}

\section*{VII. CONCLUSIONS.}

In this paper, the problem of subharmonic generation in a nonlinear acoustic
resonator is treated from the point of view of nonlinear dynamics. A linear
stability analysis reveals the possibility of time dependent solutions,
which depending on parameters can be regular or chaotic. Numerical analysis
are consistent with the theoretical predictions.

Althought the proposed model presents many analogies with similar systems
studied in nonlinear optics, such as the two-level laser or the optical
parametric oscillator, the peculiarities of acoustical resonators (specially
the non-equidistant mode distribution, and the loss mechanisms) motivates a
careful analysis. We conclude that the observability of the phenomena is
fundamentally restricted by existence of low (experimentally achievable)
threshold values, a situation which strongly depends on the resonator linear
properties, such as mode distribution and quality factors. We believe that
the proposed model could explain some experimental observations of complex
dynamical behaviour in acoustic resonators reported in the past \cite{cook89}%
.

\section*{VIII. ACKNOWLEDGEMENTS}

The authors thank Dr. H. Hobaek and Dr. Y.N. Makov for interesting
discussions on the subject. The work was financially supported by the CICYT
of the Spanish Government, under the\ project BFM2002-04369-C04-04.

\end{document}